\documentclass[english]{article}[12pt]
\pdfoutput=1
\usepackage{color}
\usepackage[latin9]{inputenc}
\usepackage{graphicx,amssymb,amsfonts,amsmath,amssymb,amscd,amstext, simplewick}
\usepackage{hyperref}
\usepackage{cite}
\usepackage{wasysym}
\usepackage{esint}
\usepackage{caption}
\usepackage{subcaption}

\newcommand{\CH}[1]{{}}

\def\be{\begin{eqnarray}}
\def\ee{\end{eqnarray}}

\def\d{{\rm d}}
\def\tr{\operatorname{tr}}

\begin{document}

\thispagestyle{empty}

\begin{flushright}
YITP-SB-15-17
\end{flushright}

\begin{center}
\vspace{1cm} { \LARGE {\bf
Estimation for Entanglement Negativity of Free Fermions}}
\vspace{1.1cm}

Christopher P.~Herzog and Yihong Wang \\
\vspace{0.8cm}
{ \it C.~N.~Yang Institute for Theoretical Physics,
Department of Physics and Astronomy \\
Stony Brook University, Stony Brook, NY  11794}

\vspace{0.8cm}

\end{center}

\begin{abstract}
\noindent
In this letter we study the negativity of one dimensional free fermions.
We derive the general form of the $\mathbb{Z}_{N}$ symmetric term
 in moments of the 
partial transposed (reduced) density matrix, which is an algebraic function of the 
end points of the system. Such a path integral turns out to be a convenient
tool for making estimations for the negativity.

\end{abstract}

\section{Introduction}

Measures of quantum entanglement have become a focus of intense research
activity at the boundaries between quantum information, quantum field
theory, condensed matter physics, general relativity and string theory
(see refs.\ \cite{Casini:2009sr,Calabrese:2009qy,EislerPeschel} for reviews).  One
key quantity, the entanglement entropy, measures the quantum entanglement
between two complementary pieces of a system in a pure state. However
the entanglement entropy is no longer a good measure of quantum entanglement
if the initial state of the system is mixed.  Negative eigenvalues in the partial transpose of the density matrix $\rho^{T_2}$
implies quantum entanglement even 
in a (bipartite) mixed state scenario
 \cite{Peres:1996dw, Horodecki:1996nc}.
This observation led to the proposal of the negativity \cite{2000JMOp...47.2151L,Vidal:2002zz},
which was later demonstrated to be a good entanglement measure \cite{PlenioPRL}.

Like the entanglement entropy, the negativity in a quantum field theory
can be computed by employing the replica trick \cite{Calabrese:2012ew, Calabrese:2012nk}.
In this setting, the negativity is the $N\rightarrow1$
limit of the partition function of an $N$-sheeted spacetime. In practice,
these partition functions can only be computed in special cases 
\cite{Calabrese:2012ew,Calabrese:2012nk}. For conformal field theories in $1+1$ dimensions,
the negativity of the single interval and the two adjacent interval cases is
determined by conformal symmetry.\footnote{%
 See ref.\ \cite{Blondeau-Fournier:2015yoa} for an extension to the massive case.
}
 Another special case where the
negativity can be determined, at least for $N > 1$, is
a massless free scalar field in 1+1 dimensions. In this case, the $N$-sheeted
partition functions are known in terms of Riemann-Siegel theta functions
although it is not known in general how to continue the result away
from integer $N > 1$ and in particular to $N =1$. Since the partial transposed reduced
density matrix is Gaussian, the negativity for a free scalar can be
checked through a lattice computation by using Wick's Theorem \cite{Calabrese:2012nk,Audenaert}.

The case of free fermions in 1+1 dimensions appears to be
more difficult than the case of free scalars however. The partial transpose of the reduced
density matrix is no longer Gaussian but a sum of two, generically non-commuting, Gaussian matrices 
\cite{EislerZimboras}:
%\CH{changed $e^{i \pi/4} \to e^{-i \pi/4}$}
\begin{align}
\rho^{T_2} = \frac{1}{\sqrt{2}} \left( e^{i \pi / 4} O_+ + e^{-i \pi / 4} O_- \right) \ .
\label{EZresult}
\end{align}
(We will define $O_\pm$ in section \ref{sec:bg}.)
This fact brings additional complication to both the lattice and field
theoretical calculations.  On the lattice side, eigenvalues of 
$(\rho^{T_2})^N$ 
 cannot be simply derived
from eigenvalues of a covariance matrix as in the Gaussian case. In a
field theory setting, one has to sum over partition functions 
with different spin structure, corresponding to different terms in the 
expansion of $(e^{i \pi /4} O_+ + e^{-i \pi /4} O_-)^N$. 
Various efforts have been made to
tame the difficulties in deriving the negativity of free fermions: On the
lattice side, algebraic simplification and numerical diagonalization
of products of these two Gaussian matrices yields the $N >1$ moments of negativity 
for the two disjoint interval case \cite{EislerZimboras, Coser:2015mta}.\footnote{%
  See also ref.\ \cite{EislerZimborastwo} for an extension to two spatial dimensions.
}
(Monte-Carlo and tensor network methods have also been used to calculate
negativity for the Ising model \cite{ising1, ising2, ising3} which, although not identical to the Dirac fermion,
is closely related.) The analytical form of such moments are derived
by evaluation of the corresponding path integrals \cite{Coser:2015eba,Coser:2015dvp}.
However in the existing results the sheet number $N$ does not appear
as a continuous variable;  it remains an open problem
 how to
take the $N\rightarrow1$ limit to get the negativity.\footnote{%
See \cite{2016arXiv160107492C} for recent progress on negativity for fermionic systems.
}

In this letter
we shall introduce a $\mathbb{Z}_{N}$-symmetric free fermion with
specific choice of spin structure. 
This fermion has several nice features that we
believe will help us explore and understand the features of free fermion
negativity.  1) The partition function explicitly reproduces the correct adjacent interval
limit.  2) The $N\rightarrow 1$ limit of the $N$ sheeted path integral can
be easily derived.  3) There exists a natural generalization to multiple interval
cases, nonzero temperature, and nonzero chemical potential. 
4) While such a partition function is not an $N^{\rm th}$ moment of $\rho^{T_2}$ 
(except in the special case $N=2$), it appears to be a useful
quantity for bounding these $N^{\rm th}$ moments including the negativity itself.
%writing upper and lower bounds for the true negativity
%and its $N$th moments.

 The rest of this letter is arranged
as follows: In section \ref{sec:bg} we review previous results.
Section \ref{sec:bosonization} contains a derivation of
the partition function for the 
$\mathbb{Z}_{N}$-symmetric free fermion system and in particular
$\tr(O_+^N)$ and $\tr[(O_+ O_-)^{N/2}]$.  
In section \ref{sec:bounds},
we discuss bounds
on the negativity and its $N^{\rm th}$ moments.
% using the partition function of the $\mathbb{Z}_{N}$-symmetric
%system. 
We conclude in section \ref{sec:final} with remarks on 
possible generalizations of our results and future directions.
An appendix contains a discussion of a two-spin system.

\section{Review of Previous Results}
\label{sec:bg}

We first review the definition of the negativity.
For a state $\left|\Psi\right\rangle $ in a quantum system with bipartite
Hilbert space ${\cal H={\cal H}}_{A}\bigotimes{\cal H}_{B}$ and density
matrix $\rho=\left|\Psi\right\rangle \left\langle \Psi\right|$, the reduced
density matrix is defined as $\rho_{A}=\tr_{B}\rho$ . If ${\cal H}_{A}$
is factored further into ${\cal H}_{A}={\cal H}_{A_{1}}\bigotimes{\cal H}_{A_{2}}$, 
one can define the partial transpose of the reduced density matrix
$\rho_{A}^{T_{2}}$ as the operator such that the following identity
holds for any $e_{i}^{\left(1\right)}$, $e_{k}^{\left(1\right)}\in{\cal H}_{A_{1}}$
and $e_{j}^{\left(2\right)}$, $e_{l}^{\left(2\right)}\in{\cal H}_{A_{2}}$:
$\left\langle e_{i}^{\left(1\right)}e_{j}^{\left(2\right)}\left|\rho_{A}^{T_{2}}\right|e_{k}^{\left(1\right)}e_{l}^{\left(2\right)}\right\rangle =\left\langle e_{i}^{\left(1\right)}e_{l}^{\left(2\right)}\left|\rho_{A}\right|e_{k}^{\left(1\right)}e_{j}^{\left(2\right)}\right\rangle $.
The logarithmic negativity is defined as the logarithm of the trace norm%
\footnote{The trace norm of a matrix $M$ is defined as the sum of its singular values: $|M| \equiv \tr\left[\left(M^{\dagger}M\right)^{1/2}\right]$.
%=\tr\left(\left(MM^{\dagger}\right)^{1/2}\right)$.  
For
Hermitian matrices, singular values are absolute values of the eigenvalues.
} 
of $\rho_{A}^{T_{2}}$. Since $\rho_{A}^{T_{2}}$ is Hermitian, its
trace norm can be written as the following limit 
%$\rho_{A}^{T_{2}}$
\be
{\cal E} \equiv \log|\rho_A^{T_2} | =\log\lim_{N_{e}\rightarrow1}\tr\left(\rho_{A}^{T_{2}}\right)^{N_{e}}
\ee
where $N_{e}$ is an even integer.  This analytic continuation suggests the utility of also defining higher moments of the partial transpose:
\be
{\cal R}(N) \equiv \tr \left[ (\rho_A^{T_2})^N \right] \ .
\ee
%The system considered in \cite{Calabrese:2012ew, Calabrese:2012nk}
We are interested in systems in one time and one spatial dimension.  We will assume a factorization of the Hilbert space corresponding to a partition of the real line with $A_{1}$ and $A_{2}$ each being the
union of a collection of disjoint intervals: $A_{1}=\cup_{i=1}^{p}\left(s_{i},t_{i}\right)$ and 
$A_{2}=\cup_{i=1}^{q}\left(u_{i},v_{i}\right)$.

In this paper, we are particularly interested in the case of free, massless fermions in 1+1 dimension with 
the continuum Hamiltonian
\be
H = \mp i  \int_{0}^L \Psi^\dagger(t,x)  \partial_x \Psi(t,x) \,  \d x   \ 
\ee
where $\{ \Psi^\dagger(t,x), \Psi(t,x') \} = \delta(x-x')$.  The sign determines whether the fermions are left moving or right moving.
We will take one copy of each to reassemble a Dirac fermion.   
It will often be convenient to consider the lattice version of this Hamiltonian as well
\be
H = \mp \frac{i}{2} \sum_j \left( \Psi^\dagger_j \Psi_{j+1} - \Psi^\dagger_{j+1} \Psi_j \right) \ ,
\ee
and anticommutation relation $\{ \Psi^\dagger_j, \Psi_k \} = \delta_{jk}$, which suffers the usual fermion doubling problem.  We choose as our vacuum the state annihilated by all of the $\Psi_j$.  

The authors of ref.\ \cite{EislerZimboras} were able to give a relatively simple expression for the negativity in the discrete case by working instead with Majorana fermions $a_{2j-1} = \frac{1}{2}( \Psi^\dagger_j + \Psi_j)$ and $a_{2j} = \frac{1}{2i}(\Psi^\dagger_j - \Psi_j)$.  
Re-indexing, we can write the reduced density matrix as a sum over words made of the $a_j$:
\be
\rho_A = \sum_{\tau} c_{\tau}\prod_{j=1}^{2n} a_j^{\tau_j}
\ee 
where $\tau_j$ is either zero or one, depending on whether the word $\tau$ contains the Majorana fermion $a_j$, and $n$ is the length of region $A$.  
Consider now instead the matrices $O_{\pm}$ constructed from $\rho_A$ by multiplying all the $a_j$ in region $A_2$ by $\pm i$:
\be
O_\pm = \sum_{\tau, \sigma} c_{\tau,\sigma} \left( \prod_{j=1}^{2n_1} a_j^{\tau_j} \right) \left( \prod_{j=2n_1+1}^{2n_1+2n_2} (\pm i a_j)^{\sigma_j} \right) \ .
\ee 
Here $n_j$ is the length of region $A_j$, and we have broken the sum into words $\tau$ involving region $A_1$ and words $\sigma$ involving region $A_2$.  
As we already described in eq.\ (\ref{EZresult}), the central result of ref.\ \cite{EislerZimboras} is that the partial transpose of the reduced density matrix can be written in terms of $O_\pm$.
%\be
%\rho_A^{T_2} = \frac{1}{\sqrt{2}} \left( e^{i \pi / 4} O_+ + e^{-i \pi / 4} O_- \right) \ .
%\ee

While the spectrum of $\rho_A$ is not simply related to the spectra of $O_\pm$, it is true that $O_+$ and $O_-$ are not only Hermitian conjugates but are also related by a similarity transformation and so have the same eigenvalue spectrum.  Consider a product of all of the Majorana fermions in $A_2$,
\be
S = i^{n_2}  \prod_{j=2n_1+1}^{2(n_1+n_2)} a_j \ ,
\ee
which squares to one, $S^2=1$.  This operator provides the similarity transformation between $O_+$ and $O_-$, i.e.\ $O_+ = S O_- S$.  This similarity transformation means, along with cyclicity of the trace, that if we have a trace over a word constructed from a product of $O_+$ and $O_-$, the trace is invariant under the swap $O_+ \leftrightarrow O_-$.  Employing this similarity transformation, the negativity for the first few even $N$ can be written thus
\be
\tr[ (\rho_A^{T_2})^2 ] &=& \tr (O_+ O_-) \ ,
\label{negtwo}
 \\
\tr[ (\rho_A^{T_2})^4 ] &=& - \frac{1}{2} \tr(O_+^4) + \tr(O_+^2 O_-^2) + \frac{1}{2} \tr[ (O_+ O_-)^2] \ ,
\label{negfour}
 \\
\tr[  (\rho_A^{T_2})^6] &=& - \frac{3}{2} \tr (O_+ O_-^5) + \frac{1}{4} \tr[(O_+ O_-)^3] + \frac{3}{4} \tr(O_+^3 O_-^3) + \frac{3}{2} 
\tr (O_+ O_- O_+^2 O_-^2)  \ .
\label{negsix}
\ee

To obtain analytic expressions for $\tr [(\rho_A^{T_2})^N ]$ from the decomposition (\ref{EZresult}) of $\rho_A^{T_2}$, a key step   \cite{Coser:2015mta} is the relation between matrix elements of $\rho_A$ and matrix elements of $O_\pm$.  Consider arbitrary coherent states $\langle \zeta(x) |$ and $| \eta(x) \rangle$ that further break up into $\langle \zeta_1(x_1), \zeta_2(x_2) | $ 
and $| \eta_1(x_1), \eta_2(x_) \rangle$ according to the decomposition of $A$ into $A_1$ and $A_2$.  Then the matrix elements of $\rho_A$ and $O_\pm$ are related via
\be
\langle \zeta(x) | O_\pm | \eta(x) \rangle = \langle \zeta_1(x_1), \pm \eta_2^*(x_2) | U_2^\dagger \rho_A U_2 | \eta_1(x_1), \mp \zeta_2^*(x_2) \rangle \ ,
\label{OptorhoA}
\ee
where $U_2$ is a unitary operator (whose precise form \cite{Coser:2015mta} does not concern us) that acts only on the part of the state in region $A_2$.

In pursuit of an analytic expression, let us move now to a path integral interpretation of $\tr [ \rho_A^N ]$ and $\tr [ (\rho_A^{T_2})^N ]$.     
The trace over $\rho_A^N$ becomes a path integral over an $N$ sheeted cover of the plane, branched over $A$.  
Now consider instead $\tr O_+^N$ given the relation (\ref{OptorhoA}). 
Performing a change of variables, we can replace $U_2$ acting on $\zeta_2^*$ and $\eta_2^*$ with $\zeta_2$ and $\eta_2$ inside the trace, and we see that $\tr (\rho_A^N)$ is related to $\tr O_+^N$ by an orientation reversal of region $A_2$.  In terms of the $N$ sheets, fixing a direction, passing through an interval in $A_1$, we move up a sheet while passing through an interval in $A_2$ we move down  a sheet.  
Indeed, the trace of any word constructed from the $O_+$ and $O_-$ has a similar path integral interpretation.

\begin{figure}[htpb]
        \centering
                    \includegraphics[width=2.5in]{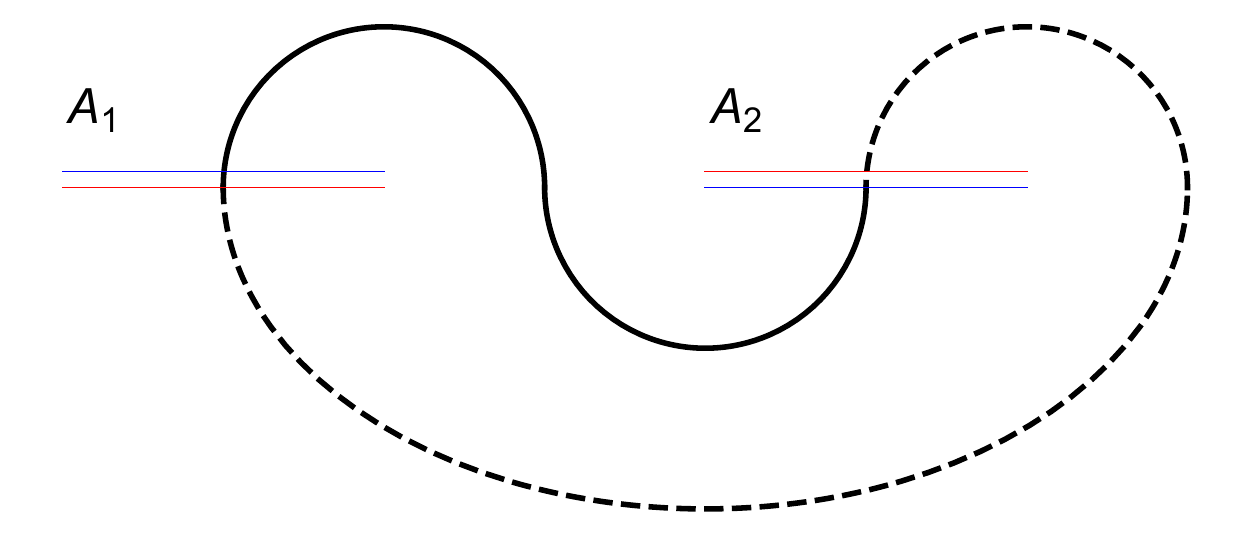}
                          \caption{
                \label{fig:fundamentalloops}   
        }
\end{figure}

Given the sign flip relation $O_- = S O_+ S$ however, replacing some of the $O_+$ by $O_-$ in the word will change the spin structure of the $N$ sheeted cover.  In particular, consider a word $\tr \left[ \prod_{i=1}^N O_{s_i} \right]$ where the $n$th and $(n+1)$th letters are both $O_+$.  Now replace the $(n+1)$th letter with $O_-$.  Any cycle passing (once) through the corresponding cut in $A_2$ between the $n$th and $(n+1)$th sheet will now pick up a minus sign compared to the situation before the replacement.  In figure \ref{fig:fundamentalloops}, we show a cycle that would pick up such a sign.

For simplicity, consider the case where $A_1$ is a single interval bounded by $s<t$ and $A_2$ a single interval bounded by $u<v$.  The trace of a word constructed from $O_\pm$, up to an undetermined over-all normalization $c_N$, can be written in terms of a Riemann-Siegel theta function \cite{Coser:2015mta}
\be
\tr \left[ \prod_{i=1}^N O_{s_i} \right]= c_N^2 \left( \frac{1-x}{(t-s) (v-u)} \right)^{2 \Delta_N} \left| \frac{\Theta[{\bf e}](\tilde \tau(x))}{\Theta(\tilde \tau(x))} \right|^2 \ , \; \; \; {\bf e} = \left( \begin{array}{c} {\bf 0} \\ {\boldsymbol\delta} \end{array} \right) \ ,
\label{twointervalword}
\ee
where ${\bf 0}$ is a vector of $N-1$ zeros and ${\boldsymbol\delta}$ is fixed by the word $ \prod_{i=1}^N O_{s_i}$.  In particular, if $s_i \neq s_{i+1}$, then $\delta_i = 1/2$ and $\delta_i = 0$ otherwise.   
The characteristic $\delta_i = 0$ is associated with having antiperiodic boundary conditions around the corresponding fundamental cycle, while the characteristic $\delta_i=1/2$ has periodic boundary conditions \cite{Coser:2015eba}.
The exponent 
\be
\Delta_N = \frac{c}{12} \left( N - \frac{1}{N} \right) 
\ee
 is the dimension of a twist operator field with $c=1$ for a Dirac fermion.  The cross ratio is defined to be
\be
x \equiv \frac{(s-t)(u-v)}{(s-u) (t-v)} \in (0,1) \ .
\ee
(The limit in which the intervals become adjacent corresponds to $x \to 1$.)  The Riemann-Siegel theta function is defined as
\be
\Theta[{\bf e}]({\bf z}|M) \equiv \sum_{{\bf m} \in {\mathbb Z}^{N-1}} e^{i \pi ({\bf m} + {\boldsymbol\epsilon})^t \cdot M \cdot ({\bf m} + {\boldsymbol\epsilon}) + 2 \pi i ({\bf m} + {\boldsymbol\epsilon})^t \cdot ({\bf z} + {\boldsymbol \delta})} \ , \; \; \; {\bf e} \equiv \left( \begin{array}{c} {\boldsymbol\epsilon} \\ {\boldsymbol\delta} 
\end {array} \right) \ ,
\label{thetafunction}
\ee
and further $\Theta({\bf z}|M) \equiv \Theta[{\bf 0}]({\bf z}|M)$.  
The $(N-1) \times (N-1)$ period matrix is then \cite{Calabrese:2012nk,Calabrese:2009ez}
\be
\tau_{i,j} = i \frac{2}{N} \sum_{k=1}^{N-1} \sin(\pi k/N) \frac{{}_2 F_1 (k/N, 1-k/N; 1; 1-x)}{{}_2 F_1 (k/N, 1-k/N; 1; x) } \cos[2 \pi (k/N)(i-j)] \ , 
\ee
and further $\tilde \tau(x) = \tau(x/(x-1))$.  
There are Riemann-Siegel theta functions that one can write down for multiple interval cases as well, but we shall not need their explicit form.

Among the words that enter in the binomial expansion of $\tr [ (\rho_A^{T_2})^N ] $, the traces
$\tr (O_+^N) = \tr (O_-^N)$ and $\tr[ (O_+ O_-)^{N/2}]$ are special.  Even in the multiple interval case, these two traces
can be expressed as rational functions of the endpoints of the intervals.  Although we have no proof in general, observationally
it seems to be true that among the words of a fixed length $\tr (O_+^N)$ is the smallest in magnitude while $\tr[(O_+ O_-)^{N/2}]$
is the largest.  These two considerations suggest the utility of trying to bound the negativity using the rational functions
$\tr(O_+)^N$ and $\tr[(O_+ O_-)^{N/2}]$, as we pursue in section \ref{sec:bounds}.

In the two interval case, it follows from the result (\ref{twointervalword}) that $\tr(O_+^N)$ and $\tr[(O_+ O_-)^N]$ are rational functions.  
That $\tr(O_+^N)$ reduces to a rational function is obvious since ${\boldsymbol \delta} = {\bf 0}$.  That $\tr[(O_+ O_-)^{N/2}]$ reduces as well follows from Thomae's formula \cite{Nakayashiki,FarkasZemel} that when $\delta_i = 1/2$ for all $i$.
\be
\left| \frac{\Theta[{\bf e}](\tilde \tau)}{\Theta(\tilde \tau)} \right|^2 = \left|1-x\right|^{-N/4} \ .
\ee
To see more generally that these words are rational functions of the endpoints, in the next section we employ 
bosonization.\footnote{%
 For an application of Thomae's formula to a multiple interval R\'enyi entropy computation, see ref.\ \cite{Coser:2013qda}.
}  

\section{Bosonization and Rationality}
\label{sec:bosonization}

Consider the normalized partition function of the free Dirac field on the $\mathbb{Z}_{N}$-curve
defined by the following set:
\be
\label{riemann}
X_{N}=\left\{ \left(z,y\right)\left|y^{N}=\prod_{i=1}^{p}\frac{z-s_{i}}{z-t_{i}}\prod_{i=1}^{q}\frac{z-v_{i}}{z-u_{i}},
\left(z,y\right)\in\mathbb{C}^{2}\right.\right\}  \ .
\ee
One can see that $X_{N}$, as the set of all points in $\mathbb{C}^{2}$
satisfying the equation in the set, has $N$ sheets corresponding
to $N$ different roots of a nonzero complex number. These $N$ copies
of ${\mathbb C}$ are cut open along intervals in $A$ on the real axis. 
As we choose the ordering $s_i < t_i$ and $u_i < v_i$,
such open
cuts are glued cyclicly if in $A_{1}$ and anti-cyclicly if in $A_{2}$.

While the Riemann surface (\ref{riemann}) has an explicit ${\mathbb Z}_N$ symmetry,
to specify a partition function, we also have to give the spin structure.  The spin
structure can generically break this symmetry, i.e.\ we can associate relative factors of minus one
to cycles that would otherwise be related by the ${\mathbb Z}_N$ shift symmetry.  
A generic word $\prod_i O_{s_i}$ will generically have a spin structure that does not
respect this symmetry.  However, a few words do, namely $\tr(O_+^N) = \tr (O_-^N)$
and $\tr[(O_+ O_-)^{N/2}]$.  The word $\tr(O_+)^N$ preserves the natural anti-periodic boundary conditions, while the word $\tr[(O_+ O_-)^{N/2}]$ associates an additional $-1$ to fundamental cycles that intersect both $A_1$ and $A_2$.

If we assume the ${\mathbb Z}_N$ symmetry is preserved by the spin structure, then the bosonization procedure is especially simple.  
Denote the partition function on $X_{N}$ by $Z[N]$.
Rather
than a path integral of a single Dirac field on $X_{N}$ in (\ref{riemann}),
$Z\left[N\right]$ can be considered as a path integral of a vector
valued Dirac field $\vec{\Psi}\left(z\right)$ on $\mathbb{C}$:
$\boldsymbol{\Psi}\left(x\right)=\left(\Psi_{1}\left(z\right), \cdots, \Psi_{N}\left(z\right)\right)$.
$\Psi_{i}\left(x\right)$
is the value of the original field $\Psi$ at coordinate $\left(z,y_{i}\right)$
on $X_{N}$. 
%This construction leads to a multi-valued vector. Therefore,
When going anti-clockwise around a branch point $w$ by a small enough
circle $C_{w}$, $\boldsymbol{\Psi}\left(x\right)$ gets multiplied
by a monodromy matrix $T\left(w\right)$.

Define the matrix 
\be
T\equiv \left(\begin{array}{ccccc}
0 & \omega\\
 &  & \omega\\
 &  & . & .\\
 &  &  & 0 & \omega\\
\omega &  &  &  & 0
\end{array}\right)\label{eq:T}
\ee
where $\omega=e^{2\pi i\frac{N-1}{N}}$.  This value of $\omega$ is
chosen so that $T$ satises the overall boundary condition $T^{N}=\left(-1\right)^{N-1}{\rm id}$ 
where
${\rm id}$ is the $N\times N$ identity matrix. The reason for
the factor $\left(-1\right)^{N-1}$ comes from considering a closed
loop that circles one of the branch points $N$ times. Such a loop
should be a trivial closed loop in the $y$ coordinate and come with
an overall factor of $-1$, standard from performing a $2\pi$ rotation
of a fermion.\footnote{%
In order to preserve an explicit ${\mathbb Z}_N$ symmetry,
we have chosen a slightly different matrix than in
ref.\ \cite{Casini:2005rm}.% 
 %is $\left(\begin{array}{ccccc}
%0 & 1\\
% &  & 1\\
 %&  & . & .\\
 %&  &  & 0 & 1\\
%\left(-1\right)^{N-1} &  &  &  & 0
%\end{array}\right)$, 
%
}

The matrix $T$ is not the only ${\mathbb Z}_N$ 
symmetric matrix satisfying $T^{N}=\left(-1\right)^{N-1}{\rm id}$.
%(\ref{eq:T}). 
A relative phase $e^{i2\pi k/N}$, $k=1,2,\ldots , N-1$,
between monodromy matrices at different branch points is also allowed.
Choose the basis of $\boldsymbol{\Psi}\left(x\right)$ so that $T\left(s_{1}\right)=T$
and take into account the constraint that $T\left(t_{i}\right)T\left(s_{i+1}\right)={\rm id}$,
$T\left(v_{i}\right)T\left(u_{i+1}\right)={\rm id} $.  Then,  the monodromy
matrices are fixed to be
\begin{align}
T\left(s_{i}\right) & =T \ ,   &  T\left(t_{i}\right) &=T^{-1}\label{eq:BDCST} \ , \\
T\left(u_{i}\right) & =\exp\left(2\pi i\left(N-k\right)/N\right)T^{-1} \ ,     & T\left(v_{i}\right) &=\exp\left(2\pi ik/N\right)T \ ,
\label{eq:BDCUV}
\end{align}
For us, $e^{2 \pi i k / N}$ represents an extra phase, in addition to the conventional anti-periodic boundary condition, when $\boldsymbol{\Psi}$ is transported around a cycle of the Riemann surface.

If we insist on the usual spin structure for fermions, that $\boldsymbol{\Psi}$ can only pick up an overall factor of $\pm 1$ around any closed cycle, then two values of $k$ are singled out, $k=0$ for all $N$ and $k=N/2$ for even $N$.  
The choice $k=0$ will produce a partition function that computes  $\tr(O_+^N)$,while the choice $k=N/2$ will produce a partition function that computes $\tr [ (O_+ O_-)^{N/2}]$.  
As we will discuss below, there are a pair of additional special choices, $k= (N \pm 1)/2$ for odd $N$, which do not have an interpretation as a $\tr[ \prod_i O_{s_i}]$, but which nevertheless have some nice properties.
For now, we will keep the dependence on $k$ arbitrary.

As introduced in refs.\ \cite{Calabrese:2012nk, Casini:2005rm, Bershadsky:1988ta, Dixon:1986qv},
a twist operator $\sigma_{R}^{k}\left(w\right)$ is defined as the
field that simulates the following monodromy behavior: 
$\vec{\Psi}\left(x\right)\sigma_{R}^{k}\left(w\right)\rightarrow\exp\left(2\pi ik/N\right)T^{R}\vec{\Psi}\left(x\right)\sigma_{R}^{k}\left(w\right)$
when $x$ is rotated counter-clockwise around $w$. Then $Z\left[N\right]$
can be expressed as a correlation function of twist operators on a single copy of ${\mathbb C}$ rather
than as a partition function on $X_N$,
\be
Z[N] \sim \left\langle\left(\prod_{i=1}^{p}\sigma_{1}^{0}\left(s_{i}\right)\sigma_{-1}^{0}\left(t_{i}\right)\prod_{j=1}^{q}\sigma_{-1}^{k}\left(u_{j}\right)\sigma_{1}^{k}\left(v_{j}\right)\right)_{{\cal AO}}\right\rangle \ .
\ee
The subscript ${\cal AO}$ means the operators are in ascending order
of coordinates. Such correlation functions can be calculated through
bosonization (see e.g.\ ref.\ \cite{Casini:2005rm}).
Diagonalization of $T$
leads to $N$ decoupled fields, $\widetilde{\Psi}_{l}$. 
Each $\widetilde{\Psi}_{l}$
is multivalued, picking up a phase $e^{-i\frac{l}{N}2\pi}$, $e^{i\frac{l}{N}2\pi}$,
$e^{i\frac{l-k}{N}2\pi}$or $e^{-i\frac{l-k}{N}2\pi}$ when rotated
counter-clockwise around $s_{i}$, $t_{i}$, $u_{i}$, or $v_{i}$ respectively.
Then one can factorize each multi-valued field $\widetilde{\Psi}_{l}$
into a gauge factor that describes this multi-valuedness and a single
valued free Dirac field: $\Psi^{l}=e^{i\int_{x_{0}}^{x}dx'^{\mu}A_{\mu}^{l}\left(x\right)}\psi^{l}\left(x\right)$.
The gauge field dependent part of the partition function contains
the branch point dependence of $Z[N]$ and is moreover straightforward
to evaluate. With the notation 
%$q_{l}\left(R,k\right)$ in 
%Ref. {[}\cite{key-17}{]}:
 \cite{Bershadsky:1988ta},
\be
q_{l}\left(R,k\right)\equiv
\frac{1-N}{2N}+\left\{ \frac{lR+k+\left(N-1\right)/2}{N}\right\} \ , 
\ee
where the curly braces denote the fractional part of a number and
$l\in\ell=\left\{ -\frac{N-1}{2},-\frac{N-1}{2}+1,...,\frac{N-1}{2}\right\} $,
the gauge field $A_{\mu}^{l}\left(x\right)$ satisfies the contour
integrals
\begin{align}
\oint_{C_{s_{i}}}dx^{\mu}A_{\mu}^{l}\left(x\right)  &=-\frac{2\pi l}{N} \ ,  & \oint_{C_{s_{i}}}dx^{\mu}A_{\mu}^{l}\left(x\right) &=\frac{2\pi l}{N}
\label{eq:phasest} \ , \\
\oint_{C_{u_{i}}}dx^{\mu}A_{\mu}^{l}\left(x\right)   &=2\pi q_{l}\left(1,N-k\right) \ ,  & \oint_{C_{v_{i}}}dx^{\mu}A_{\mu}^{l}\left(x\right) &=2\pi q_{l}\left(-1,k\right) \ .
\label{eq:phaseuv}
\end{align}
The Lagrangian density%
\footnote{Our conventions for the Clifford algebra are that $\left\{ \gamma^{\mu},\gamma^{\nu}\right\} =2\delta^{\mu\nu}$.  For
example, we could choose $\gamma^{x}=\sigma^{3}$ and $\gamma^{t}=\sigma^{1}$%
}
 in terms of $\psi^{l}\left(x\right)$ becomes ${\cal L}=\sum_{l=1}^{N}\bar{\text{\ensuremath{\psi}}}^{l}\gamma^{\mu}\left(\partial_{\mu}+iA_{\mu}^{l}\right)\psi^{l}$.
From eqs.\ (\ref{eq:phasest}) and (\ref{eq:phaseuv}) and Green's
theorem we have:
\begin{equation}
\epsilon^{\mu\nu}\partial_{\nu}A_{\mu}^{l}\left(x\right)=2\pi\sum_{i=1}^{p}\sum_{j=1}^{q}\left[\frac{l}{N}\left(\delta\left(x-s_{i}\right)-\delta\left(x-t_{i}\right)\right)-q_{l}\left(1,N-k\right)\left(\delta\left(x-u_{i}\right)-\delta\left(x-v_{i}\right)\right)\right] \ .
\end{equation}
Since the $\psi^{l}$'s are decoupled, the partition function becomes
a product of expectation values of operators that depend on the gauge
field $A_{\mu}$:
\be
{\cal T}\left[N\right] \equiv
\frac{Z\left[N\right]}{(Z\left[1\right])^N}=\prod_{l\in\ell}\langle e^{i\int A_{\mu}^{l}j_{l}^{\mu}d^{2}x}\rangle \ , 
\ee
where $j_{l}^{\mu}$ is the Dirac current $\bar{\psi^{l}}\gamma^{\mu}\psi^{l}$.
After bosonization, it becomes $j_{l}^{\mu}=\frac{1}{2\pi}\epsilon^{\mu\nu}\partial_{\nu}\phi^{l}$.
Then ${\cal T}\left[N\right]$ can be written as a correlation function
of free boson vertex operators $V_{e}\left(w\right)=e^{-i\frac{e}{2}\phi_{l}\left(w\right)}$,
\begin{equation}
\prod_{l=-\frac{N-1}{2}}^{\frac{N-1}{2}}\langle e^{i\int A_{\mu}^{l}j_{l}^{\mu}d^{2}x}\rangle=
\left\langle\prod_{i=1}^{p}\prod_{j=1}^{q}V_{2l/N}\left(s_{i}\right)V_{-2l/N}\left(t_{i}\right)V_{2q_{l}\left(-1,k\right)}\left(u_{j}\right)V_{2q_{l}\left(1,N-k\right)}\left(v_{j}\right)\right\rangle \ .
\end{equation}

To evaluate the correlation function of twist operators, we use 
\begin{equation}
\left\langle\prod_{l_{i}=1}^{m}V_{e_{i}}\left(w_{i}\right)\right\rangle=\prod_{i\neq j}\left|w_{i}-w_{j}\right|^{-e_{i}e_{j}}\epsilon^{-m}\label{eq:vcr}
\end{equation}
where $\epsilon$ is a UV cut-off to take into account the effect of coincident
points in the correlation function.  
We also need the sums 
\be
\sum_{l=-\frac{N-1}{2}}^{\frac{N-1}{2}}\frac{l^{2}}{N^{2}}&=&\frac{N^{2}-1}{12N} \ ,\\
\sum_{l=-\frac{N-1}{2}}^{\frac{N-1}{2}}\frac{lq_{l}\left(1,N-k\right)}{N^{2}}&=&\frac{N^{2}-1}{12N}-\frac{\left(N-k\right)k}{2N} \ .
\ee
to  get an explicit expression for ${\cal T}\left[N\right]$. 

To shorten the expressions, we adopt the following notation:
$\left\{ s_{i}\right\} =S;\left\{ t_{i}\right\} =T;\left\{ u_{i}\right\} =U;\left\{ v_{i}\right\} =V$
along with
\be
\left[Y,Z\right]=\left|\prod_{y\in Y,z\in Z}\left(y-z\right)\right| \ , \; \; \;
\left[Y,Y\right]=\left|\prod_{y_1,y_2\in Y,y_1\neq y_2}\left(y_1-y_2\right)\right| \ .
\ee
Then ${\cal T}\left[N\right]$ can be written as:
\begin{equation}
{\cal T}[N] = L^{- \frac{N^2-1}{6N}} X^{ \frac{N^2-1}{6 N} - \frac{(N-k)k}{N}} \ ,
\end{equation}
where we have defined
\begin{equation}
L \equiv \frac{\left[S,T\right]\left[U,V\right]}{\left[S,S\right]\left[T,T\right]\left[U,U\right]\left[V,V\right]\epsilon^{p+q}} \ , \; \; \;
X \equiv \frac{\left[S,V\right]\left[T,U\right]}{\left[S,U\right]\left[T,V\right]} \ .
\end{equation}
%\begin{equation}
%{\cal T}\left[N\right]=\left(\frac{\left[S,T\right]\left[U,V\right]}{\left[S,S\right]\left[T,T\right]\left[U,U\right]\left[V,V\right]\epsilon^{p+q}}\right)^{-\frac{N^{2}-1}{6N}}\left(\frac{\left[S,V\right]\left[T,U\right]}{\left[S,U\right]\left[T,V\right]}\right)^{2\left(\frac{N^{2}-1}{12N}-\frac{\left(N-k\right)k}{2N}\right)}\label{eq:TN}
%\end{equation}
Fixing the appropriate spin structures, we claim then that 
\be
\tr(O_+^N) = \tr(O_-^N) &=&
 \left( \frac{L}{X} \right)^{- \frac{N^2-1}{6N}} \ , \\
%c_N^2 \left(\frac{\left[S,T\right]\left[U,V\right] \left[ S,U  \right] \left[T,V \right]
%}{
%\left[S,S\right]\left[T,T\right]\left[U,U\right]\left[V,V\right]
%\left[S,V \right] \left[ T,U \right]
%\epsilon^{p+q}}\right)^{-\frac{N^{2}-1}{6N}
%} \ ,
%\\
%
\tr[(O_+ O_-)^{N/2}] &=&
\left( \frac{L}{X} \right)^{- \frac{N^2-1}{6N}}  X^{-N/4} \ .
% L^{- \frac{N^2-1}{6N} } X^{- \frac{N^2+2}{12 N}} \ .
%c_N^2 \left(\frac{\left[S,T\right]\left[U,V\right]}{\left[S,S\right]\left[T,T\right]\left[U,U\right]\left[V,V\right]
%\epsilon^{p+q}}\right)^{-\frac{N^{2}-1}{6N}}\left(\frac{\left[S,V\right]\left[T,U\right]}
%{\left[S,U\right]\left[T,V\right]}\right)^{- \frac{N^2+2}{12N}} \ .
\ee
Comparing with the two interval case (\ref{twointervalword}), we can absorb $c_N$ into the $\epsilon$
dependence of $L$. 
A nice feature of these expressions is that it is straightforward to take the $N \to 1$ limit.

\subsection{Adjacent Limits}

Let us consider adjacent limits of the two-interval negativity.  We call the single-interval negativity the case when $s=v$ and $t=u$, and there is only one length scale, say $l = t-s$.  We call the two-adjacent-interval negativity the case where $t=u$ and we have two length scales, $l_1 = t-s$ and $l_2 = v-u$.  
The single-interval and two-adjacent-interval
negativities are given by a two point function and a three point function
of twist fields respectively.  They are therefore fully determined by conformal symmetry \cite{Calabrese:2012ew,Calabrese:2012nk}:
\begin{eqnarray}
{\cal R}\left(N_{o}\right)\sim l^{-\frac{N_{o}^{2}-1}{6N}} \ ,  & \quad & {\cal R}\left(N_{e}\right)\sim l^{-\frac{N_{e}^{2}-4}{6N}}\label{eq:2pt} \ , \\
{\cal R}\left(N_{o}\right)\sim\left(l_{1}l_{2}\left(l_{1}+l_{2}\right)\right)^{-\frac{N_{o}^{2}-1}{12N}} \ , & \quad & {\cal R}\left(N_{e}\right)\sim\left(l_{1}l_{2}\right)^{-\frac{N_e^{2}-4}{12N}}\left(l_{1}+l_{2}\right)^{-\frac{N_e^{2}+2}{12N}}\label{eq:3pt} \ .
\end{eqnarray}
While $\tr(O_+^N)$ simply vanishes in these coincident limits, we claim that $\tr[(O_+ O_-)^{N/2}]$ reproduces 
${\mathcal R}(N_e)$ for even $N$, in both the single-interval and two-adjacent-interval cases.  
This agreement provokes the question is there a choice of $k$ for odd $N$ for which ${\mathcal T}[N]$ has the correct adjacent interval limits?  The answer is yes.  If we choose $k = (N_o \pm 1)/2$, then 
\be
{\cal T}\left[N_o\right]=
L^{-\frac{N_o^2-1}{6N_o}} X^{-\frac{N_o^2-1}{12N_o}} \ ,   
%\left(\frac{\left[S,T\right]\left[U,V\right]}{\left[S,S\right]\left[T,T\right]
%\left[U,U\right]\left[V,V\right]\epsilon^{p+q}}\right)^{-\frac{N_o^{2}-1}
%{6N_o}}\left(\frac{\left[S,V\right]\left[T,U\right]}{\left[S,U\right]\left[T,V\right]}\right)^{- \frac{N_o^2-1}{12N_o}}
\ee
and this expression reproduces ${\mathcal R}(N_o)$ in the adjacent interval limits.

To see why the values  $k = N_e/2$ and $k = (N_o \pm 1)/2$ are singled out, we consider the merging of twist operators
 $\sigma_1^k(w_i) \sigma_1^0(w_{i+1}) \to \sigma_2^k(w_i)$.  The corresponding
constraint on the correlation function is
\begin{equation}
\lim_{w_{i+1}\rightarrow w_{i}}\left\langle \sigma_{R_{1}}^{k_{1}}\left(w_{1}\right)\cdots\sigma_{1}^{k}\left(w_{i}\right)\sigma_{1}^{0}\left(w_{i+1}\right)\cdots\right\rangle \left|w_{i}-w_{i+1}\right|^{
%e_i e_{i+1}
-\gamma_{i\left(i+1\right)}
}=\left\langle \sigma_{R_{1}}^{k_{1}}\left(w_{1}\right)\cdots\sigma_{2}^{k}\left(w_{i}\right)\cdots\right\rangle 
\end{equation}
along with a corresponding constraint from considering $\sigma_{-1}^{0}\left(w_{i}\right)\sigma_{-1}^{k}\left(w_{i+1}\right)$.
We have defined 
\be
\gamma_{ij} \equiv \sum_{l\in\ell}q_{l}\left(R_{i},k_{i}\right)q_{l}\left(R_{j},k_{j}\right) \ .
\ee
These constraints can only be satisfied if the following identities holds for all $l 
\in \ell$:
\begin{equation}
q_{l}\left(-2,k\right)=q_{l}\left(-1,0\right)+q_{l}\left(-1,k\right) \ , \qquad q_{l}\left(2,k\right)=q_{l}\left(1,0\right)+q_{l}\left(1,k\right) \ .\label{eq:cdk}
\end{equation}
The $k$ values $(N_o-1)/2$, $N_e/2$ and $(N_o+1)/2$ are the only solutions.

\section{Bounds on the Negativity}
\label{sec:bounds}

We discuss three types of bounds on ${\cal R}\left(N\right)$
in the following subsections.  The first, which follows from a triangle inequality on the Schatten $p$-norm, is an upper bound on the moments of the partially transposed density matrix.  The second two are conjectural.  We are able to demonstrate these conjectured bounds only for small $N > 1$.

The Schatten $p$-norm, 
defined as 
\be
\left\Vert M\right\Vert _{p}\equiv \left(\tr\left(\left(M^{\dagger}M\right)^{p/2}\right)\right)^{1/p} \ , \quad   p\in\left[1,\infty\right) \ ,
\ee
is a generalization of the trace norm.
Indeed, the Schatten 1-norm is the trace norm. 

Because 
%$\left({\cal R}\left[N\right]\right)^{1/N}$
$\tr [ (\rho_A^{T_2})^N]^{1/N}$
is the Schatten $N$-norm of $\rho_A^{T_{2}}$, for all even $N$ we
have by the triangle inequality that 
\begin{equation}
%{\cal E}\left[N\right]=
\tr[(\rho_A^{T_2})^N] =  
\left(\left\Vert \rho_{A}^{T_{2}}\right\Vert _{N}\right)^{N}\leq
2^{-N/2}
\left(\left\Vert 
%\frac{1-i}{2}
e^{i \pi /4}O_{+}
\right\Vert _{N}+\left\Vert 
%\frac{1+i}{2}
e^{-i \pi/4}O_{-}
\right\Vert _{N}\right)^{N}=
2^{N/2} \tr[ (O_+ O_-)^{N/2} ] \ .
%2\tr\left(\left(\frac{1}{2}O_{+}O_{-}\right)^{\frac{N}{2}}\right)=\left(\sqrt{2}\right)^{N}{\cal T}\left[N\right]
\label{eq:tib}
\end{equation}
The $N\rightarrow1$ limit of (\ref{eq:tib}) leads to an upper bound
on the negativity in terms of $\tr[ (O_+ O_-)^{1/2} ]$ 
%${\cal E}$ in terms of ${\cal T}$
\be
{\cal R}(1)=\left\Vert \rho_{A}^{T_{2}}\right\Vert _{T}\leq\left\Vert \frac{1+i}{2}O_{+}\right\Vert _{T}+\left\Vert \frac{1-i}{2}O_{-}\right\Vert _{T}
 = \sqrt{2} \tr [ (O_+ O_-)^{1/2} ] = \sqrt{2} X^{-1/4} \ .
%=2\tr\left(\left(\frac{1}{2}O_{+}O_{-}\right)^{\frac{1}{2}}\right)=\sqrt{2}{\cal T}=\sqrt{2}X^{-\frac{1}{4}}
\label{triangleupper}
\ee
We have thus established that $\tr[(O_+ O_-)^{N/2}]$ provides a rigorous upper bound on the negativity and its $N$th moments, for free fermions.

\subsection{Conjecture 1: Bounds from Word Order}

%Here are some notations we use: 
%\CH{changed a sign}
%$O_{\pm}=\alpha \pm \beta$ and  
%%$O_{-}=\alpha+\beta$; 
%$\rho_{A}^{T_{2}}=\alpha+i\beta$ where
%$\alpha$ is a hermitian block diagonal matrix $\left(\begin{array}{cc}
%A & 0\\
%0 & C
%\end{array}\right)$ and $\beta=\left(\begin{array}{cc}
%0 & -B\\
%B^{\dagger} & 0
%\end{array}\right)$ %
%\footnote{Such a representation of $O_{\pm}$ is allowed because $O_{-}=SO_{+}S$
%and $O_{+}^{\dagger}=O_{-}$%
%}. The other two useful notions are $\gamma\equiv BS=\left(\begin{array}{cc}
%0 & B\\
%B^{\dagger} & 0
%\end{array}\right)$ by $\gamma$and $\eta_{\pm}\equiv\left(\begin{array}{cc}
%A & \pm B\\
%\pm B^{\dagger} & -C
%\end{array}\right)$. Note that $\eta_{\pm}$ are hermitian and $\eta_{+}=SO_{+}=O_{-}S\quad\eta_{-}=O_{+}S=SO_{-}$.
%Also we denote the cross ratio $\frac{\left[S,V\right]\left[T,U\right]}{\left[S,U\right]\left[T,V\right]}$
%by $X$ and $\frac{\left[S,T\right]\left[U,V\right]}{\left[S,S\right]\left[T,T\right]\left[U,U\right]\left[V,V\right]\epsilon^{p+q}}$
%by $L$ 

%\subsection*{Bounds for ${\cal E}\left[N_{e}\right]$ by Order of Words}

 As we discussed briefly above, for words of a fixed, even length,
we conjecture that $\tr (O_\pm^N)$ is the smallest and $\tr [ (O_+ O_-)^{N/2} ]$ is the largest
among the traces.  
In the notation of the previous section, we expect that the trace of an arbitrary word
$O_+^{n_1} O_-^{n_2} \cdots$ of length $N$ is bounded above and below by
\be
\tr (O_+^N) = \tr[ (O_+ O_-)^{N/2}] X^{N/4} \leq  \tr(O_+^{n_1} O_-^{n_2} \cdots) \leq \tr[ (O_+ O_-)^{N/2}]   \ .
\ee

\begin{figure}[htpb]
        \centering
        \begin{subfigure}[b]{0.5\textwidth}
                \includegraphics[width=\textwidth]{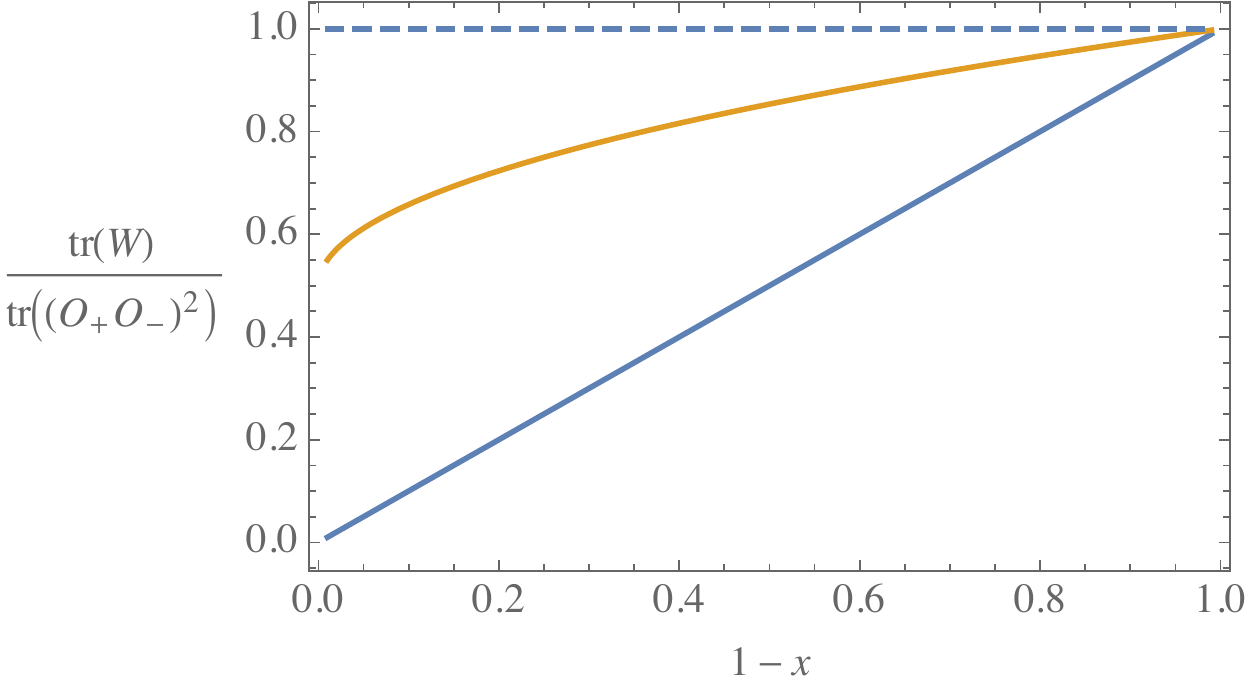}
                \caption{ $N=4$}
                \label{fig:woa}
        \end{subfigure}%
        ~
        \begin{subfigure}[b]{0.5\textwidth}
                \includegraphics[width=\textwidth]{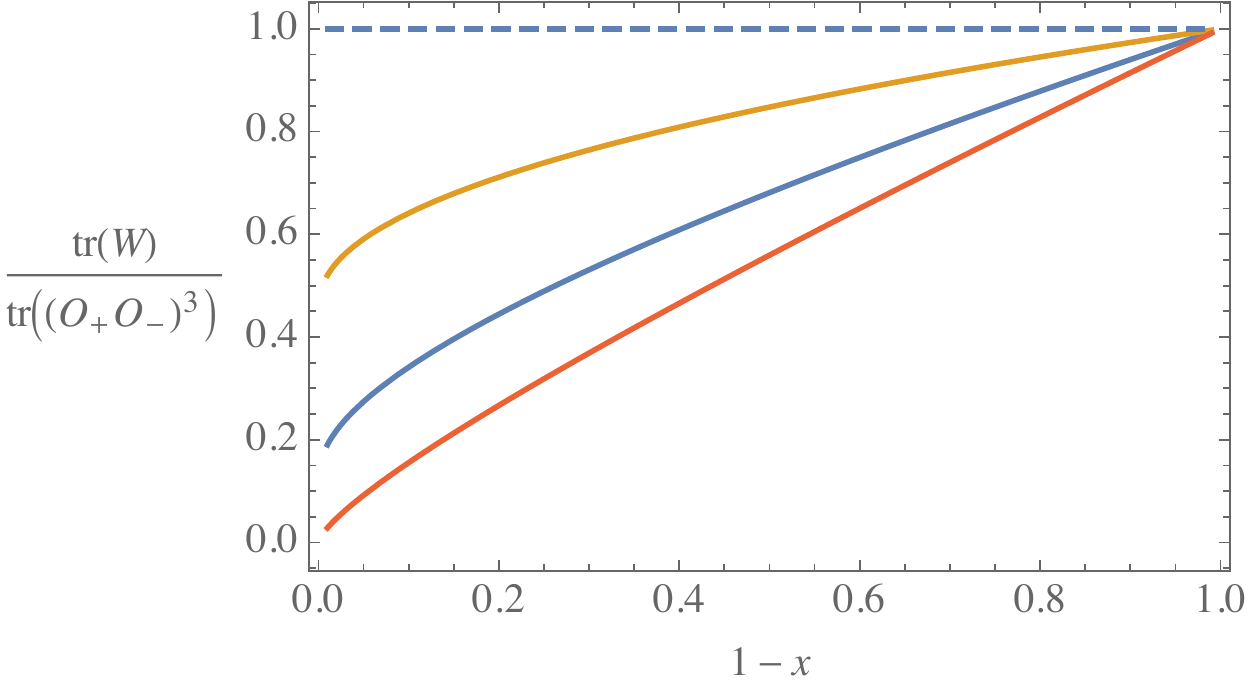}
                \caption{ $N=6$}
                \label{fig:wob}
        \end{subfigure}%
     
               \caption {Plots of ratios of traces of words versus the four point ratio $1-x$ for the two disjoint interval system. 
               In the $N=4$ case, we compare $\tr(O_+^4)$ and $\tr(O_+^2 O_-^2)$ to $\tr[(O_+ O_-)^2]$. The lowest curve
               is the ratio of  $\tr(O_+^4)$  to  $\tr[(O_+ O_-)^2]$.
               In the $N=6$, we compare $\tr(O_+^5 O_-)$, $\tr(O_+ O_- O_+^2 O_-^2)$ and $\tr(O_+^3 O_-^3)$ to $\tr[ (O_+ O_-)^3]$.  The curve at the bottom corresponds to the ratio of $\tr(O_+^5 O_-)$ to $\tr[ (O_+ O_-)^3]$ and establishes that $\tr(O_+^5 O_-)$ is the smallest among the words that appears in the negativity. 
               The dashed line is included as a guide to the eye.
               %we first establish that $\tr[(O_+ O_-)^6]$ is largest.  Then we establish that $
               %               is $N$th moment of the partial transposed reduced density matrix. 
               \label{fig:wordorder}}
\end{figure}

We can refine this conjecture on word order further.  Define $s = |n_+ - n_-|$ to be the difference between the number of times $n_+$ that $O_+$ appears in a word and the times $n_-$ that $O_-$ appears in a word.  For two words $W_1$ and $W_2$, we conjecture that if $s(W_1) > s(W_2)$, then $\tr(W_1) < \tr (W_2)$.  Indeed, we have checked this conjecture in the two interval case for small $N$, using the explicit representation of these traces in terms of Riemann-Siegel theta functions.  See figure \ref{fig:wordorder}.  

Given this refined conjecture on word order, we can obtain upper and lower bounds on the negativity.  For an upper bound, we first consider a binomial expansion of $\tr[(\rho_A^{T_2})^{N}]$.  Each word in the expansion will come with a coefficient proportional to $e^{\pm s i \pi / 4}$.  We take advantage of the $O_+ \leftrightarrow O_-$ symmetry of the words to restrict to words with $n_+ \geq n_-$
and replace $e^{\pm s i \pi / 4}$ with $\cos (\pi s /4)$.   In so doing, we eliminate all the words of charge $s = 2$ mod 4. 
Indeed, as we move from left to right in a row toward the middle of Pascal's triangle, $s$ decreases by two at each step and the coefficients follow a repeating pattern 
$(+,0,-, 0)$.  Consider all the terms that appear in the binomial expansion with a positive sign such that $s \neq 0$.   
We replace every such word with charge $s$ with a word of charge $s-4$ and hence larger trace.  Because the number of words grows as the charge decreases, we will still have a net negative contribution from words of charge $s-4$.  We then replace all the traces of words with negative coefficient by the yet smaller trace $\tr(O_+)^N$.  For the words of charge $s=0$, we simply replace all of them by the larger $\tr[(O_+ O_-)^{N/2}]$.  At the end of this procedure, we find the following upper bound
\be
\tr[ (\rho_A^{T_2})^{N}] \leq \left[1 - \frac{1}{2^{N/2}} {N \choose \frac{N}{2} } \right] \tr(O_+^N) + \frac{1}{2^{N/2}}  {N \choose \frac{N}{2} } \tr[(O_+ O_-)^{N/2}] \ .
\ee 
The coefficient of the first term is a sum over the coefficients in the binomial expansion with the words of charge $s=0$ removed.
  
In the large $N$ limit, the right hand side of this expression approaches
\be
\sqrt{\frac{2^{N+1}}{\pi N}} \left( \tr [(O_+ O_-)^{N/2}] - \tr(O_+^N) \right) \ ,
\ee
which appears to be a somewhat more stringent condition than our rigorous upper bound (\ref{eq:tib}).

We can obtain a lower bound in a similar fashion, reversing the procedure.  We consider all the terms in the binomial expansion of 
$\tr[(\rho_A^{T_2})^{N}]$ that appear with negative coefficient.  We replace every such word with charge $s$ by a word of charge $s-4$.  All the traces will then have positive coefficient.  Next, except for $\tr [(O_+ O_-)^{N/2}]$ itself,  we replace all the traces of words with the smaller $\tr(O_+)^N$.  In this case, we find the lower bound
\be
\left(1 - \frac{1}{2^{N/2 - 1}} \right) \tr (O_+^N) + \frac{1}{2^{N/2 - 1}} \tr[(O_+ O_-)^{N/2}] \leq \tr[ (\rho_A^{T_2})^{N}]  \ .
\ee
Here the coefficient of the first term is a sum over the binomial coefficients with only the word $\tr[(O_+ O_-)^{N/2}]$ removed.
In comparison with the conjecture we discuss next, this lower bound is not particularly stringent in the large $N$ limit.  

We can establish these bounds rigorously only for small $N$.  Note that for $N=2$, the upper and lower bound reduce to the known equality (\ref{negtwo}).  For $N=4$ and $N=6$, we obtain the constraints
\begin{align}
\frac{1}{2}(1 + X) \tr[ (O_+ O_-)^2]  & \leq \tr[(\rho_A^{T_2})^4] \leq \frac{1}{2} ( 3 - X) \tr[ (O_+ O_-)^2] \ , 
\label{eq:b4} 
\\ 
%\tr[(\rho_A^{T_2})^6] &\leq \frac{1}{2} (5 - 3 X^{3/2}) \tr[(O_+ O_-)^3] \ .
\frac{1}{4} (1+  3X^{3/2} ) \tr[(O_+ O_-)^3]  & \leq \tr[(\rho_A^{T_2})^6] \leq \frac{1}{2} (5 - 3 X^{3/2}) \tr[(O_+ O_-)^3] \ .
\label{eq:b6}
\end{align}
Indeed, in the two interval case, using the explicit representation of the negativity in terms of Riemann-Siegel theta functions, we can verify that these bounds are indeed satisfied.  See the insets in figure \ref{fig:TN1}. 

For $N=4$, we can do better and prove the inequalities in general.
That $\tr[|O_+^2 - O_-^2|^2] \geq 0$ implies that $\tr (O_+^4) \leq \tr (O_+^2 O_-^2)$.  
Similarly, that $\tr[|O_+ O_- - O_- O_+|^2] \geq 0$ implies that $\tr (O_+^2 O_-^2) \leq \tr [(O_+ O_-)^2]$
and the desired inequalities on $\tr[ (\rho_A^{T_2})^4]$ follows directly.\footnote{%  
 Alternately, one can employ von Neumann's trace inequality.
}
It is tempting to apply these inequalities to the case $N=1$.

\subsection*{Conjecture 2: A Lower Bound from Extremization}

\begin{figure}
  \centering
        \begin{subfigure}[b]{0.5\textwidth}
                \includegraphics[width=\textwidth]{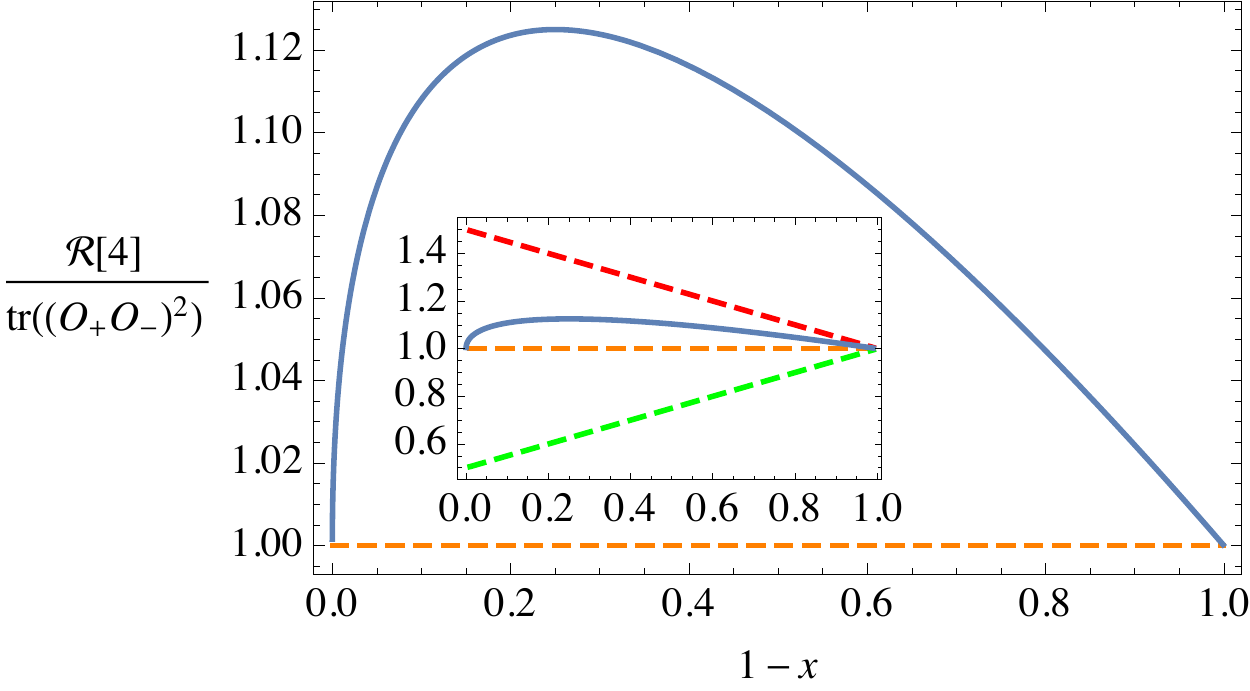}
                \caption{ $N=4$}
                \label{fig:1a}
        \end{subfigure}%
        ~ 
        \begin{subfigure}[b]{0.5\textwidth}
                \includegraphics[width=\textwidth]{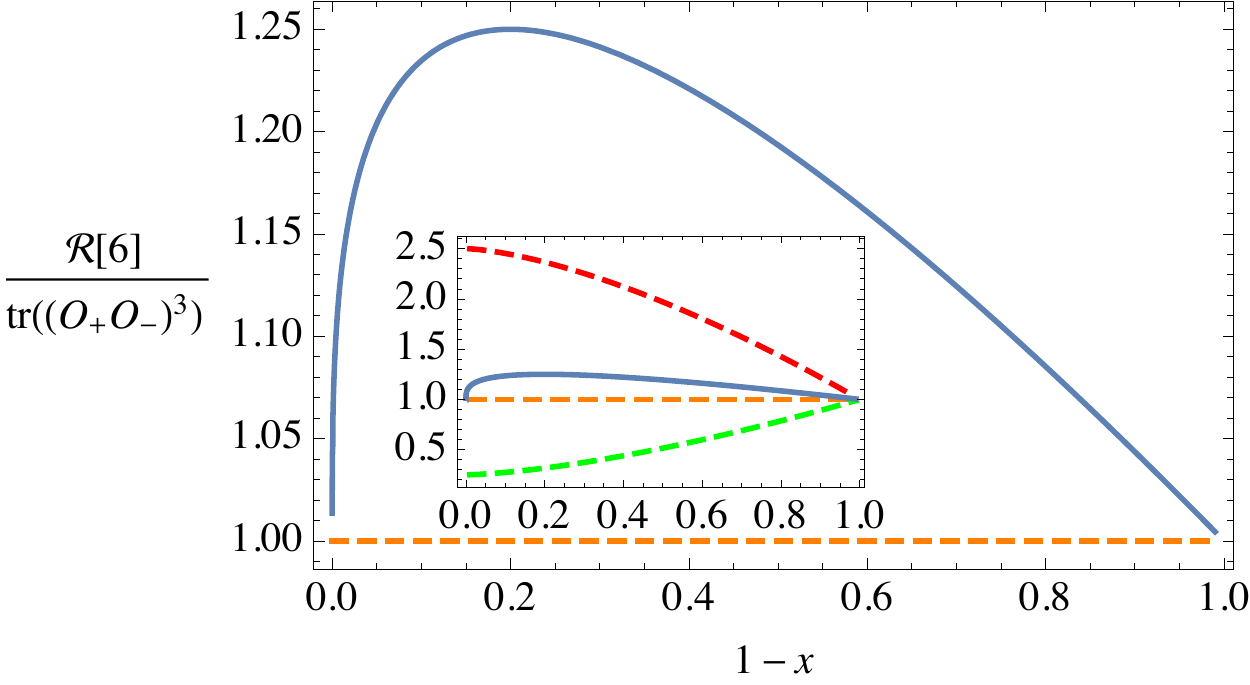}
                \caption{ $N=6$}
                \label{fig:1b}
        \end{subfigure}%
        \caption{
        Proposed bounds on the negativity.  The solid blue line is $\tr[(\rho_A^{T_2})^N] / \tr[(O_+ O_-)^{N/2}]$.  The dashed line is the constant function 1.  In the insets, the upper and lower bounds (\ref{eq:b4}) and (\ref{eq:b6}) are included.  The horizontal axis is the 
four point ratio $1-x$. 
%\CH{Do we know the $x\to 0$ limit?} 
}
\label{fig:TN1}
\end{figure}

The plot of the two disjoint interval system suggests another possible
type of lower bound on ${\cal R}\left(N\right)$.  
At least for $N=2$, 4 and 6, and conjecturally for all even $N$, 
we find that 
\be
\tr [ (O_+ O_-)^{N/2}] \leq
{\cal R}(N) = \tr[(\rho_{A}^{T_{2}})^{N}] \ .
\label{extremallower}
\ee
%{\cal R}\left[N\right]$. 
%The matrix representation of this inequality
% is $\tr\left(\left(O_{+}O_{-}\right)^{N/2}\right)\leq \tr\left(\left(\rho_{A}^{T_{2}}\right)^{N}\right)$.
%Interestingly, analytically continuing to $N=3$ and 5, the inequality seems to hold for odd $N$ as well. 
Figure \ref{fig:TN1} is a comparison of the ratio $\tr[(\rho_A^{T_2})^N] / \tr[(O_+ O_-)^{N/2}]$ as a function of the four point ratio $x$ to the constant 
function one.  We consider $N=4$ and $N=6$ for the two interval case only.
For $N=2$, the inequality is saturated given (\ref{negtwo}).  Given the saturation, we further conjecture
that the negativity itself is bounded above, 
\be
{\mathcal E} = \log\left(| \rho_A^{T_2}| \right)\leq \log\left(\tr[(O_+ O_-)^{1/2}]\right) \ ,
\label{extremalupper}
\ee  
further tightening the triangle inequality (\ref{triangleupper}).
In the appendix, we compute the $N$th moments 
$\tr[(\rho^{T_2})^N]$ and $\tr[(O_+ O_-)^{N/2}]$ explicitly for a two-spin system in a Gaussian state.  We are able to show that 
the bounds (\ref{extremallower}) and (\ref{extremalupper}) 
are satisfied in this simple case.\footnote{%
 As a consistency check, note that this upper bound is in general larger than the lower bound (46) of  ref.\ \cite{EislerZimboras} (under the assumption that $\tr_{e/o}\left(O_\pm\right)$ is real) and becomes identical if the even part of $O_+$ is Hermitian and the odd part of $O_-$ is  anti-Hermitian with negative imaginary part.
 }
%
%$\tr[(O_+ O_-)^{N/2}] \leq \tr[(\rho^{T_2})^N]$ for $N>2$ while for $N=1$, ${\cal R} \leq \tr[(O_+ O_-)^{1/2}]$. 

We can try to put more structure behind this conjecture.  
We begin by introducing some notation.
Recalling that $O_+^\dagger = O_-$ and that $O_+ = S O_- S$, we can assume
without loss of generality the following block structure for $O_\pm$:
\begin{equation}
O_\pm = \left( \begin{array}{cc}
A & \pm B \\
\mp B^\dagger & C 
\end{array}
\right)
\end{equation}
where $A$ and $C$ are Hermitian.  It will be useful in what follows to consider
\be
\alpha \equiv \left( \begin{array}{cc}
A & 0 \\
0 & C 
\end{array}
\right) \ , \; \; \;
\beta \equiv \left( \begin{array}{cc}
0 & B \\
-B^\dagger & 0 
\end{array}
\right) \ ,
\ee
such that $O_\pm = \alpha \pm \beta$ and, from (\ref{EZresult}), $\rho_A^{T_2} = \alpha + i \beta$. 
Finally, we introduce
\be
\gamma\equiv S \beta=\left(\begin{array}{cc}
0 & B\\
B^{\dagger} & 0
\end{array}\right)
 \ , \; \; \;
 \eta_{\pm}\equiv\left(\begin{array}{cc}
A & \pm B\\
\pm B^{\dagger} & -C
\end{array}\right) \ .
 \ee
 Note that the $\eta_{\pm}$ are Hermitian and that $\eta_{+}=SO_{+}=O_{-}S$ while $\eta_{-}=O_{+}S=SO_{-}$.

Define the function
\be
f_N(\theta) \equiv \tr [ ((\alpha  + e^{i \theta} \beta) (\alpha - e^{-i \theta}\beta))^{N/2} ] \ .
\ee
From this definition, it follows that $f_N(\frac{\pi}{2}) = \tr[(\rho_A^{T_2})^N ]$ and 
$f_N(0) = \tr[(O_+ O_-)^{N/2}]$.  
This function has a few other useful properties.  It is periodic, with period $2 \pi$:
$f_N(\theta) = f_N(\theta + 2\pi)$.  
It also has two reflection symmetries.  The first,
$f_N(\theta) = f_N(\pi-\theta)$, follows from cyclicity of the trace:
\begin{eqnarray*}
f_N(\pi - \theta) &=&  \tr [ ((\alpha  - e^{-i \theta} \beta) (\alpha + e^{i \theta}\beta))^{N/2} ] \\
&=&  \tr [ ( (\alpha + e^{i \theta} \beta)(\alpha  - e^{-i \theta} \beta) )^{N/2} ] \\
&=& f_N(\theta) \ .
\end{eqnarray*}
The second, $f_N(\theta) = f_{N}(-\theta)$, is more subtle. Consider expanding out the product of matrices inside the trace. 
A generic term in the product will involve $n_+$ factors of $e^{i \theta} \beta$ and $n_-$ factors $-e^{-i \theta} \beta$.  
If $n_+ = n_-$, then the $\theta$ dependence drops out, and such terms are irrelevant for the argument that follows.  Let us therefore assume $n_+ \neq n_-$.  
Because $\beta$
is off diagonal, any term that contributes to the trace must have an even number of factors of $\beta$.  
Thus either $n_+$ and $n_-$ are both odd or both even.  
For every such term, there will also be a term with $n_+$ factors of $-e^{-i \theta} \beta$ and $n_-$ factors of $e^{i \theta} \beta$.  
  This second term will always have the same sign and coefficient as the first and the same cyclic ordering of operators.  Thus, we can re-express the $\theta$ dependence of the combined terms as $\cos( (n_+ - n_-) \theta)$, which is an even function of $\theta$.
  
  The two reflection symmetries, $f(\theta) = f(-\theta)$ and $f(\theta) = f(\pi-\theta)$ along with periodicity imply that 
  $f(\pi/2) = f(3\pi/2)$ are extrema of $f(\theta)$ as are $f(0) = f(\pi)$.  
  If we can show that these four extrema are the only extrema in the domain $0 \leq \theta < 2 \pi$, and that $f(\pi/2)$ is a local maximum (or alternatively that $f(0)$ is a local minimum), then our conjecture is proven since $f(\theta)$ is a smooth bounded function on this domain.  

For even $N$, the difference between the first few ${\cal R}\left(N\right)$
and $\tr[(O_{+}O_{-})^{N/2}]$
can be written in terms of $\alpha$ and $\gamma$: 
\begin{eqnarray}
{\cal R}\left(2\right)-{\cal T}\left[2\right] & = & 0 \ , \label{eq:d2}\\
{\cal R}\left(4\right)-{\cal T}\left[4\right] & = & 
4 \tr \left( (\alpha \gamma)^2 \right)
\ , \label{eq:d4} \\
%2\tr\left(\left(\alpha\gamma\right)^{2}+\left(\gamma\alpha\right)^{2}\right) \ , \label{eq:d4}\\
{\cal R}\left(6\right)-{\cal T}\left[6\right] & = & 6 \tr\left(\left(\alpha^{2}+\gamma^{2}\right)\left(\left(\alpha\gamma\right)^{2}+\left(\gamma\alpha\right)^{2}\right)\right) \ .\label{eq:d6}
\end{eqnarray}
A sufficient condition for $\tr[(O_{+}O_{-})^{N/2}]\leq{\cal R}\left(N\right)$
to hold for $N=4$ and $N=6$ is that $\left(\alpha\gamma\right)^{2}+\left(\gamma\alpha\right)^{2}$
be positive definite.

\section{Comments and Future Directions}
\label{sec:final}

While a determination of the negativity ${\mathcal E}$ for massless free fermions in 1+1 dimensions
remains an open problem, we have argued in this paper that $\tr[(O_+ O_-)^{N/2}]$ and $\tr(O_+^N)$, which have simple closed form
expressions for all real $N$, 
can be used to bound ${\mathcal E}$ as well as higher moments of $\rho_A^{T_2}$.  
One of our main results is that 
\be
{\mathcal E} \leq  \log\left(\sqrt{2} \tr[(O_+ O_-)^{1/2}]\right)  \ ,
\ee
which follows from the triangle inequality.  Part of our Conjecture 2 is that the bound can be tightened by removing the $\sqrt{2}$.  
Also, in the appendix, we demonstrated this tighter upper bound for a two-spin system in a Gaussian state.  

For $N>2$, we have both upper and lower bounds on the moments of  $\rho_A^{T_2}$.  In their strongest form, our conjectures state that
\be
\tr[(O_+ O_-)^{N/2}] \leq \tr[(\rho_A^{T_2})^N] \leq \left[ 1 - \frac{1}{2^{N/2}} {N \choose \frac{N}{2}} \right] \tr(O_+)^N  
+ \frac{1}{2^{N/2}} {N \choose \frac{N}{2}} \tr[(O_+ O_-)^{N/2}] \ .
\ee
Using the triangle inequality, we were also able to argue rigorously for a somewhat weaker upper bound (\ref{eq:tib}).

An advantage of working with $\tr[(O_+ O_-)^{N/2}]$ and $\tr(O_+^N)$ instead of with $\tr[(\rho_A^{T_2})^N]$ is that 
they are much simpler quantities.  
In the paper, we discussed how to compute the multiple interval case on the plane.  It is straightforward to consider the torus instead, i.e.\ finite volume and nonzero temperature.\footnote{%
 See ref.\ \cite{Calabrese:2014yza} for a discussion of subtleties associated with thermal effects on negativity.
}
One can even introduce a chemical potential.  These generalizations require the use of the appropriate torus correlation function in place of eq.\ (\ref{eq:vcr}).  See for example refs.\ \cite{Ogawa:2011bz,Herzog:2013py}.

There are many interesting
questions that could be asked regarding $\tr[(O_+ O_-)^{N/2}]$. 
What can we deduce about the eigenvalues
of $\left(O_{+}O_{-}\right)^{1/2}$ from the relation $\tr[(O_{+}O_{-})^{N/2}]=L^{-\frac{N^{2}-1}{6N}}X^{-\frac{N^{2}+2}{12}}$?
Can we prove the two conjectures involving $\tr[(O_+ O_-)^{N/2}]$ discussed in the text?
Among all such open questions, the most
important and intriguing one is whether we can construct both an 
upper bound and lower bound for ${\cal R}\left(N\right)$ using $\tr[(O_+ O_-)^{N/2}]$
that have the same $N\rightarrow1$ limit.  If so, then we can extract
the value of the negativity ${\cal E}$ from these bounds.

\section*{Acknowledgments}
We would like to thank H.~Casini, D.~Park, M.~Ro\v{c}ek, T.~Hartman, V.~Korepin, and Z.~Zimboras for discussion.
We thank E.~Tonni and Z.~Zimboras for comments on the manuscript.
This work was supported in part by the National Science
Foundation under Grant No. PHY13-16617.

\appendix

\section{Two Bit System}
\label{sec:twobit}

Consider a two spin system in a Gaussian state with density matrix
\be
\rho  = \exp\left( \sum_{i=1}^2 M_{ij} c_i^\dagger c_j \right)\ ,
\ee
 where $M$ is a $2 \times 2$ Hermitian matrix and 
$c_j$ and $c_j^\dagger$ satisfy the usual anti-commutation relations, $\{ c_i, c_j^\dagger \} = \delta_{ij}$. 
In the basis $(1, c_2^\dagger, c_1^\dagger, c_1^\dagger c_2^\dagger) |0 \rangle$, such a density matrix takes
the explicit form
\be
\rho =
\frac{1}{1 + \tr e^M + \det e^M} 
\left(
\begin{array}{ccc}
1 & 0 & 0 \\
0 & e^M & 0 \\
0 & 0 & \det e^M
\end{array}
\right)
=
 \frac{1}{1+x+y+w} \left(
\begin{array}{cccc}
1 & 0 & 0 & 0 \\
0 & x & z & 0 \\
0 & \bar z & y & 0 \\
0 & 0 & 0 & w
\end{array}
\right)
\ee
where $x \geq 0$ and $y \geq 0$ while $z \in {\mathbb C}$.   Note that $w = xy - |z|^2 \geq 0$.  

The usual partial transpose of this density matrix with respect to the second spin is
\be
\rho^{\widetilde T_2} = 
\frac{1}{1+x+y+w} \left(
\begin{array}{cccc}
1 & 0 & 0 & z \\
0 & x & 0 & 0 \\
0 & 0 & y & 0 \\
\bar z & 0 & 0 & w 
\end{array}
\right) \ .
\ee
However, defining Majorana fermions $a_{2j-1} = \frac{1}{2}(c_j^\dagger + c_j)$ and $a_{2j} = \frac{1}{2i}(c_j^\dagger - c_j)$, this naive partial transpose is related to the one in the body of the paper by a similarity transformation:
$\rho^{T_2} = (\sigma_x \otimes I) (\rho^{\widetilde T_2})^T(\sigma_x \otimes I)$. More explicitly,
\be
\rho^{ T_2} = 
\frac{1}{1+x+y+w} \left(
\begin{array}{cccc}
y & 0 & 0 & 0 \\
0 & w & z & 0 \\
0 & \bar z & 1 & 0 \\
0 & 0 & 0 & x 
\end{array}
\right) \ .
\ee
Both will thus have the same spectrum.  In particular,
we find the eigenvalues
\be
\frac{1}{1+x+y+w} \left(x, y, \frac{1}{2} \left(1 + w \pm \sqrt{1-6w + w^2 + 4 xy}\right) \right) \ .
\ee
A sufficient condition for this density matrix to possess quantum entanglement is a negative eigenvalue.  We thus require $2 w < xy$.  

In the body of the paper, we also introduced the matrices $O_\pm$, which for this simple system take the explicit form
\be
O_\pm =
\frac{1}{1+x+y+w} \left(
\begin{array}{cccc}
y & 0 & 0 & 0 \\
0 & w & \pm i z & 0 \\
0 & \pm i \bar z & 1 & 0 \\
0 & 0 & 0 & x 
\end{array}
\right) \ .
\ee
As we did in the body of the paper in a more complicated case, we would like to compare the $\tr[(\rho^{T_2})^{2N}]$ with $\tr[(O_+ O_-)^N]$.  We need first the eigenvalues of $(\rho^{T_2})^2$ and $O_+ O_-$.
%,
%\be
%\frac{1}{(1+x + y + w)^2} \left( x^2, y^2, \frac{1}{2} \left( (1-w)^2 + 2xy \pm (1+w) \sqrt{1-6w + w^2 + 4 xy} \right) \right) \ ,
%\ee
%and the eigenvalues of $O_+ O_-$,
%\be
%\frac{1}{(1+x+y+w)^2} \left( x^2 , y^2 , \frac{1}{2} \left( (1-w)^2 +2xy\pm (1-w) \sqrt{(1-w)^2 + 4 xy}\right) \right) \ .
%\ee
We find that
\be
\tr[(\rho^{T_2})^{N}] - \tr[(O_+ O_-)^{N/2}] = \frac{4^n}{(1+x+y+w)^{N}} \bigl[
(A+ 1+ w)^{N} + (A-1-w)^{N}  \nonumber \\
- (B+1-w)^{N} - (B-1+w)^{N} 
\bigr]
\ee
where we have defined $A^2 \equiv (1-w)^2 -4w + 4 xy$ and $B^2 \equiv (1-w)^2  + 4 xy$.  We have used the fact that for $2w < xy$, $A > w+1$.  Note that the right hand side vanishes when $N=2$ as expected.  It also vanishes when $w=0$ and (provided $N$ is even) when $w= xy$.  
For $N=1$, the difference is negative and proportional to $A - B$, indicating that $\tr[(O_+ O_-)^{1/2}]$ is an upper bound for the negativity.  
Meanwhile for $N>2$, the right hand side is always positive over the region $0<2w<xy$, indicating that 
$\tr[(O_+ O_-)^{N/2}]$ is a lower bound on $\tr[(\rho^{T_2})^{N/2}]$. We prove this last statement below.

\subsection*{Proof of the lower bound}

We can make the further redefinitions
\be
A+1+w &=& R \cos \alpha \ , \\
A-1-w &=& R \sin \alpha \  , \\
\max\{B+1-w, B-1+w \} &=& R \cos \beta \ , \\
\min \{B+1-w, B-1+w\} &=& R \sin \beta \ ,
\ee
where $\alpha, \beta \in (0, \frac{\pi}{4})$ and $R = \sqrt{A^2 + (1+w)^2} = \sqrt{B^2 + (1-w)^2}$.  
Recalling the $N=1$ case, 
because $\cos \theta + \sin \theta$ is an increasing function in the domain $(0, \frac{\pi}{4} )$, the fact that
$A<B$ implies that $\alpha < \beta$.  We need then to establish that for $n>2$ that the following difference is positive:
\be
\tr[(\rho^{T_2})^N] - \tr[(O_+ O_-)^{N/2}] = \frac{(4 R)^N}{(1+w+y+w)^N} \left[ \cos^N \alpha + \sin^N \alpha - \cos^N\beta - \sin^N \beta \right] \ .
\ee
For $N>2$ in the domain $\theta \in (0, \frac{\pi}{4} )$, $\sin^N \theta + \cos^N \theta$ is a decreasing function:
\be
\frac{d}{d\theta} \left( \cos^N \theta + \sin^N \theta \right) = \frac{N}{2} \sin 2 \theta \left( \sin^{N-2} \theta - \cos^{N-2} \theta \right) < 0 \ .
\ee
Since $0 < \alpha < \beta < \frac{\pi}{4}$, it follows then that
\be
\cos^N \alpha + \sin^N \alpha - \cos^N \beta - \sin^N \beta > 0 \ , 
\ee
and the difference in question, $\tr[(\rho^{T_2})^N] - \tr[(O_+ O_-)^{N/2}]  > 0$, is positive.


\newpage
\begin{thebibliography}{99}


%1 
%\cite{Casini:2009sr}
\bibitem{Casini:2009sr} 
  H.~Casini and M.~Huerta,
  ``Entanglement entropy in free quantum field theory,''
  J.\ Phys.\ A {\bf 42}, 504007 (2009)
  [arXiv:0905.2562 [hep-th]].
  %%CITATION = ARXIV:0905.2562;%%
  %142 citations counted in INSPIRE as of 22 May 2015

%2
%\cite{Calabrese:2009qy}
\bibitem{Calabrese:2009qy} 
  P.~Calabrese and J.~Cardy,
  ``Entanglement entropy and conformal field theory,''
  J.\ Phys.\ A {\bf 42}, 504005 (2009)
  [arXiv:0905.4013 [cond-mat.stat-mech]].
  %%CITATION = ARXIV:0905.4013;%%
  %186 citations counted in INSPIRE as of 04 Jun 2015
  
  %3
   \bibitem{EislerPeschel}
 I.~Peschel and V.~Eisler, ``Reduced density matrices and entanglement entropy in free lattice models,''
 J.\ Phys.\  A  {\bf 42}, 504003 (2009)
 [arXiv:0906.1663 [cond-mat]].

%4
%\cite{Peres:1996dw}
\bibitem{Peres:1996dw} 
  A.~Peres,
  ``Separability criterion for density matrices,''
  Phys.\ Rev.\ Lett.\  {\bf 77}, 1413 (1996)
  [quant-ph/9604005].
  %%CITATION = QUANT-PH/9604005;%%
  %186 citations counted in INSPIRE as of 22 May 2015

%5
%\cite{Horodecki:1996nc}
\bibitem{Horodecki:1996nc} 
  M.~Horodecki, P.~Horodecki and R.~Horodecki,
  ``On the necessary and sufficient conditions for separability of mixed quantum states,''
  Phys.\ Lett.\ A {\bf 223}, 1 (1996)
  [quant-ph/9605038].
  %%CITATION = QUANT-PH/9605038;%%
  %111 citations counted in INSPIRE as of 22 May 2015


 \bibitem{2000JMOp...47.2151L} 
 J.~Lee,  M.~S.~Kim, Y.~J.~Park, and S.~Lee, 
 ``Partial teleportation of entanglement in a noisy environment,''
 J.\ of Mod.\ Opt., {\bf 47}, 2151 (2000).

%6
%\cite{Vidal:2002zz}
\bibitem{Vidal:2002zz} 
  G.~Vidal and R.~F.~Werner,
  ``Computable measure of entanglement,''
  Phys.\ Rev.\ A {\bf 65}, 032314 (2002).
  %%CITATION = PHRVA,A65,032314;%%
  %138 citations counted in INSPIRE as of 22 May 2015
 
 %7
  \bibitem{PlenioPRL}
  M.~B.~Plenio, ``Logarithmic Negativity: A Full Entanglement Monotone
  That is not Convex,''
  Phys.\ Rev.\ Lett.\ {\bf 95}, 090503 (2005).

%8
 %\cite{Calabrese:2012ew}
\bibitem{Calabrese:2012ew} 
  P.~Calabrese, J.~Cardy and E.~Tonni,
  ``Entanglement negativity in quantum field theory,''
  Phys.\ Rev.\ Lett.\  {\bf 109}, 130502 (2012)
  [arXiv:1206.3092 [cond-mat.stat-mech]].
  %%CITATION = ARXIV:1206.3092;%%
  %24 citations counted in INSPIRE as of 22 May 2015
  
%9
%\cite{Calabrese:2012nk}
\bibitem{Calabrese:2012nk} 
  P.~Calabrese, J.~Cardy and E.~Tonni,
  ``Entanglement negativity in extended systems: A field theoretical approach,''
  J.\ Stat.\ Mech.\  {\bf 1302}, P02008 (2013)
  [arXiv:1210.5359 [cond-mat.stat-mech]].
  %%CITATION = ARXIV:1210.5359;%%
  %16 citations counted in INSPIRE as of 22 May 2015

%10
%\cite{Blondeau-Fournier:2015yoa}
\bibitem{Blondeau-Fournier:2015yoa} 
  O.~Blondeau-Fournier, O.~A.~Castro-Alvaredo and B.~Doyon,
  ``Universal scaling of the logarithmic negativity in massive quantum field theory,''
  arXiv:1508.04026 [hep-th].
  %%CITATION = ARXIV:1508.04026;%%
  %1 citations counted in INSPIRE as of 02 Jan 2016

%11
\bibitem{Audenaert}
K.~Audenaert, J.~Eisert, M.~B.~Plenio and R.~F.~Werner, 
``Entanglement Properties of the Harmonic Chain,''
 Phys.\ Rev.\ A {\bf 66}, 042327 (2002).


%12
\bibitem{EislerZimboras}
V.~Eisler and Z.~Zimboras,
``On the partial transpose of fermionic Gaussian states,''
New J.\ Phys.\ {\bf 16}, 123020 (2014)
[arXiv:1502.01369 [cond-mat.stat-mech]].

%13
%\cite{Coser:2015mta}
\bibitem{Coser:2015mta} 
  A.~Coser, E.~Tonni and P.~Calabrese,
  ``Partial transpose of two disjoint blocks in XY spin chains,''
  arXiv:1503.09114 [cond-mat.stat-mech].
  %%CITATION = ARXIV:1503.09114;%%



%14
\bibitem{ising1}
H. Wichterich, J. Molina-Vilaplana, and S. Bose,
``Scaling of entanglement between separated blocks in spin chains at criticality,''
Phys.\ Rev.\ A {\bf 80}, 010304(R) (2009)
[arXiv:0811.1285 [quant-ph]].



%15
\bibitem{ising2}
V.~Alba,
``Entanglement negativity and conformal field theory: a Monte Carlo study,''
J.\ Stat.\ Mech.\ P05013 (2013)
[arXiv:1302.1110 [cond-mat.stat-mech]].

%16
\bibitem{ising3}
 P.~Calabrese,  L.~Tagliacozzo, and E.~Tonni,
``Entanglement negativity in the critical Ising chain,''
J.\ Stat.\ Mech., P05002 (2013) 
 [arXiv:1302.1113 [cond-mat.stat-mech]]. 

 %17
  %\cite{Coser:2015eba} 
  \bibitem{Coser:2015eba} A.~Coser, E.~Tonni and P.~Calabrese, 
  ``Towards entanglement negativity of two disjoint intervals for a one dimensional free fermion,'' 
  arXiv:1508.00811 [cond-mat.stat-mech].
   %%CITATION = ARXIV:1508.00811;%% %1 citations counted in INSPIRE as of 24 sept. 2015 

%\cite{Coser:2015dvp}
\bibitem{Coser:2015dvp} 
  A.~Coser, E.~Tonni and P.~Calabrese,
  ``Spin structures and entanglement of two disjoint intervals in conformal field theories,''
  arXiv:1511.08328 [cond-mat.stat-mech].
  %%CITATION = ARXIV:1511.08328;%%

\bibitem{EislerZimborastwo}
V.~Eisler and Z.~Zimboras, ``Entanglement negativity in two-dimensional free lattice models,'' 
arXiv:1511.08819 [cond-mat.stat-mech].
%%CITATION = ARXIV:1511.08819;%%

%30
%\cite{Calabrese:2009ez}
\bibitem{Calabrese:2009ez} 
  P.~Calabrese, J.~Cardy and E.~Tonni,
  ``Entanglement entropy of two disjoint intervals in conformal field theory,''
  J.\ Stat.\ Mech.\  {\bf 0911}, P11001 (2009)
  [arXiv:0905.2069 [hep-th]].
  %%CITATION = ARXIV:0905.2069;%%
  %61 citations counted in INSPIRE as of 22 Jun 2015


 %21
  \bibitem{Nakayashiki}
A.~Nakayashiki, ``On the Thomae Formula for ${\mathbb Z}_N$ Curves,''
Publ.\ RIMS, {\bf 33} 987 (1997).

%22
\bibitem{FarkasZemel}
H.~M.~Farkas and S.~Zemel, 
``Generalizations of Thomae's
Formula for $Z_{n}$ Curves,''
Dev.\ Math.\ {\bf 21}, 1 (2011).

%23
%\cite{Coser:2013qda}
\bibitem{Coser:2013qda} 
  A.~Coser, L.~Tagliacozzo and E.~Tonni,
  ``On R\'enyi entropies of disjoint intervals in conformal field theory,''
  J.\ Stat.\ Mech.\  {\bf 2014}, P01008 (2014)
  [arXiv:1309.2189 [hep-th]].
  %%CITATION = ARXIV:1309.2189;%%
  %7 citations counted in INSPIRE as of 22 Jun 2015
  



%18
%\cite{Casini:2005rm}
\bibitem{Casini:2005rm} 
  H.~Casini, C.~D.~Fosco and M.~Huerta,
  ``Entanglement and alpha entropies for a massive Dirac field in two dimensions,''
  J.\ Stat.\ Mech.\  {\bf 0507}, P07007 (2005)
  [cond-mat/0505563].
  %%CITATION = COND-MAT/0505563;%%
  %51 citations counted in INSPIRE as of 22 May 2015


%20
%\cite{Bershadsky:1988ta}
\bibitem{Bershadsky:1988ta} 
  M.~Bershadsky and A.~Radul,
  ``Fermionic Fields On Z(n) Curves,''
  Commun.\ Math.\ Phys.\  {\bf 116}, 689 (1988).
  %%CITATION = CMPHA,116,689;%%
  %8 citations counted in INSPIRE as of 22 May 2015
  


%27
%\cite{Dixon:1986qv}
\bibitem{Dixon:1986qv} 
  L.~J.~Dixon, D.~Friedan, E.~J.~Martinec and S.~H.~Shenker,
  ``The Conformal Field Theory of Orbifolds,''
  Nucl.\ Phys.\ B {\bf 282}, 13 (1987).
  %%CITATION = NUPHA,B282,13;%%
  %762 citations counted in INSPIRE as of 22 May 2015

 
   
%19   
%\cite{Calabrese:2014yza}
\bibitem{Calabrese:2014yza} 
  P.~Calabrese, J.~Cardy and E.~Tonni,
  ``Finite temperature entanglement negativity in conformal field theory,''
  J.\ Phys.\ A {\bf 48}, no. 1, 015006 (2015)
  [arXiv:1408.3043 [cond-mat.stat-mech]].
  %%CITATION = ARXIV:1408.3043;%%
  %9 citations counted in INSPIRE as of 22 May 2015


  %25
  %\cite{Ogawa:2011bz}
\bibitem{Ogawa:2011bz} 
  N.~Ogawa, T.~Takayanagi and T.~Ugajin,
  ``Holographic Fermi Surfaces and Entanglement Entropy,''
  JHEP {\bf 1201}, 125 (2012)
  [arXiv:1111.1023 [hep-th]].
  %%CITATION = ARXIV:1111.1023;%%
  %150 citations counted in INSPIRE as of 24 May 2015

  
%24
%\cite{Herzog:2013py}
\bibitem{Herzog:2013py} 
  C.~P.~Herzog and T.~Nishioka,
  ``Entanglement Entropy of a Massive Fermion on a Torus,''
  JHEP {\bf 1303}, 077 (2013)
  [arXiv:1301.0336 [hep-th]].
  %%CITATION = ARXIV:1301.0336;%%
  %13 citations counted in INSPIRE as of 22 May 2015
  
\bibitem
%[Chang \& Wen(2016)]
{2016arXiv160107492C} 
P.-Y.~Chang and X.~Wen, 
  ``Entanglement negativity in free-fermion systems: an overlap matrix approach,"
arXiv:1601.07492 [cond-mat.stat-mech].

 
%  
%  
% 
%   
% 
%
%%26
%%\cite{Klich:2015ina}
%\bibitem{Klich:2015ina} 
%  I.~Klich, D.~Vaman and G.~Wong,
%  ``Entanglement Hamiltonians for chiral fermions with zero modes,''
%  arXiv:1501.00482 [cond-mat.stat-mech].
%  %%CITATION = ARXIV:1501.00482;%%
%  
%
%
%%28
%%\cite{DiFrancesco:1997nk}
%\bibitem{DiFrancesco:1997nk} 
%  P.~Di Francesco, P.~Mathieu and D.~Senechal,
%  ``Conformal Field Theory,''
%  New York, USA: Springer (1997) 890p.
%  %%CITATION = INSPIRE-454643;%%
%  %48 citations counted in INSPIRE as of 22 May 2015
%
%
%%29
%%\cite{Wong:2013gua}
%\bibitem{Wong:2013gua} 
%  G.~Wong, I.~Klich, L.~A.~Pando Zayas and D.~Vaman,
%  ``Entanglement Temperature and Entanglement Entropy of Excited States,''
%  JHEP {\bf 1312}, 020 (2013)
%  [arXiv:1305.3291 [hep-th]].
%  %%CITATION = ARXIV:1305.3291;%%
%  %45 citations counted in INSPIRE as of 22 May 2015
%
%
%%30
%%\cite{Calabrese:2009ez}
%\bibitem{Calabrese:2009ez} 
%  P.~Calabrese, J.~Cardy and E.~Tonni,
%  ``Entanglement entropy of two disjoint intervals in conformal field theory,''
%  J.\ Stat.\ Mech.\  {\bf 0911}, P11001 (2009)
%  [arXiv:0905.2069 [hep-th]].
%  %%CITATION = ARXIV:0905.2069;%%
%  %61 citations counted in INSPIRE as of 22 Jun 2015
%
%%31
%%\cite{Calabrese:2010he}
%\bibitem{Calabrese:2010he} 
%  P.~Calabrese, J.~Cardy and E.~Tonni,
%  ``Entanglement entropy of two disjoint intervals in conformal field theory II,''
%  J.\ Stat.\ Mech.\  {\bf 1101}, P01021 (2011)
%  [arXiv:1011.5482 [hep-th]].
%  %%CITATION = ARXIV:1011.5482;%%
%  %45 citations counted in INSPIRE as of 23 May 2015
%  
%    
%   %32
%   \bibitem{Bendoukha}Bendoukha, Berrabah, and Hafida Bendahmane, ``Inequalities between the sum of powers and the exponential of sum of positive and commuting selfadjoint operators," Archivum Mathematicum 47.4, 257-262 (2011).
%
%


\end{thebibliography}
\end{document}